\documentclass[pre,aps,eqsecnum,twocolumn]{revtex4}
\usepackage[centertags]{amsmath}
\usepackage{amssymb}
\usepackage{amsthm}

\begin{document}
\title{Quantum escape kinetics over a fluctuating barrier}

\author{Pulak Kumar Ghosh$^2$, Debashis Barik$^2$, Bidhan Chandra Bag$^1$ and
Deb~Shankar~Ray$^2$ {\footnote {e-mail address:
pcdsr@mahendra.iacs.res.in}}}

\affiliation{Indian Association for the Cultivation of Science,
Jadavpur, Kolkata 700 032, India}

\begin{abstract}
The escape rate of a particle over a fluctuating barrier in a double
well potential exhibits resonance at an optimum value of correlation
time of fluctuation. This has been shown to be important in several
variants of kinetic model of chemical reactions . We extend the
analysis of this phenomenon of resonant activation to quantum domain
to show how quantization significantly enhances resonant activation
at low temperature due to tunneling.
\end{abstract}

\maketitle

\section{Introduction}
The enhancement of small periodic signal by noise in a nonlinear
system has been a theme\cite{ben} of topical interest over more than
two decades. Ever since the observation of this phenomenon of
stochastic resonance in varied theoretical and experimental
contexts, a number of noise-induced resonance effects have been
reported \cite{mcn,roy,luc1,jul,rei,lin}. A prototypical effect of
this kind,
 known as resonant
activation discovered by Doering and Godua \cite{dor}, concerns a
resonance effect in the escape rate of a particle over a fluctuating
barrier in a bistable potential, where the resonance
\cite{Bier,van,brey,shep,rein,gam,han,ros,mar}
 can be achieved by
observing the variation of mean first passage time as function of
flipping rate of fluctuation of the barrier height. In the simplest
possible term the phenomenon can be realized in a model with a
linear barrier with a slope which fluctuates between two values. The
phenomenon has triggered a lot of theoretical activity around the
Markovian and non-Markovian variants of kinetic models for chemical
reactions \cite{Bier,van,brey,rein} and
  several related issues and been experimentally observed by
Mantegna and Spagnolo \cite{ros}. This fluctuation of potential is
also important in various problems of chemical physics, notably in
molecular dissociation dynamics \cite{mad}, protein folding
\cite{wan} among others. The purpose of this paper is to extend the
analysis of resonant activation to quantum domain. Since at low
temperature thermal activation is accompanied by tunneling, the
question that naturally arises is how noise-induced resonance
effects manifest themselves in a quantum system. Specifically our
object here is two-fold: First, we intend to understand the
counterintuitive role of external noise in bringing out the
resonance behaviour in a bistable quantum system where the mean
escape time is varied as a function of correlation time of noise.
Second, it is worthwhile to examine the nature of this resonance in
presence of generic quantum effects like tunneling at low
temperature and allow ourselves a fair comparison with the classical
results. In what follows we carry out a theoretical and a numerical
study of quantum stochastic dynamics of a bistable system with a
barrier height fluctuating due to an external Ornstein-Uhlenbeck
noise and show how the mean escape time is profoundly influenced by
the statistical properties of the noise particularly the correlation
time in exhibiting the quantum resonance activation.

\section{Quantum stochastic dynamics}

\subsection{General aspects}
To derive quantum Langevin equation from a microscopic picture
consider the well-known standard system-reservoir model with
following form of the Hamiltonian\cite{zwa}

\begin{equation}\label{1.1}
\hat{H}=\frac{\hat{p}^2}{2 m}+V(\hat{x},t)+\sum_{j=1}^N \left\{
\frac{\hat{p}^2_j}{2}+\frac{1}{2} \kappa_j (\hat{q}_j-\hat{x})^2
\right\}
\end{equation}

Here $\hat{x}$ and $\hat{p}$ are the coordinate and momentum
operators of the particle and $\{\hat{q}_j, \hat{p}_j\}$ are the
set of coordinate and momentum operators for the reservoir
oscillators coupled linearly through the coupling constants
${\kappa}_j (j=1,2,...)$. The potential $V(\hat{x},t)$ is due to
the external force field for the Brownian particle. The coordinate
and momentum operators follow the usual commutation rules
$\{\hat{x}, \hat{p}\}=i\hbar$ and $\{\hat{q}_i,
\hat{p}_j\}=i\hbar{\delta}_{ij}$. Eliminating the bath degrees of
freedom in the usual way we obtain the operator Langevin equation
for the particle

\begin{equation}\label{2.2}
m \ddot{\hat{x}}+\int^{t}_0
dt'\gamma(t-t')\dot{\hat{x}}(t')+V'(\hat{x},t) = \hat{F}(t)
\end{equation}

 where the noise operator $\hat{F}(t)$
and the memory kernel $\gamma(t)$ are given by

\begin{equation}\label{2.3}
\hat{F}(t) =
\sum_j\left[\{\hat{q}_j(0)-\hat{x}(0)\}\kappa_j\cos\omega_jt
+\kappa_j^{1/2}\hat{p}_j(0) \sin\omega_jt\right]
\end{equation}

and

\begin{equation}\label{2.4}
\gamma(t) =\sum_j\kappa_j\cos\omega_jt
\end{equation}

respectively  with $\kappa_j=\omega_j^2$. On the basis of quantum
mechanical average $ \langle...\rangle $ over the bath modes with
coherent states and over the system mode with an arbitrary state
Eq.(2.2) can be cast into the form of the generalized quantum
Langevin equation \cite{db1,dbb1,dbb2,db2,db3,db4,d41,db5}.

\begin{equation}\label{2.5}
m \ddot{x} + \int_0^{t}dt'\gamma(t-t') \dot{x}(t')+V'(x,t)= f(t) +
Q(x ,\langle\delta\hat{x}^n\rangle)
\end{equation}

where the quantum mechanical mean value of the position operator
$\langle\hat{x}\rangle =x$. Here the quantum dispersion term $Q$
to the potential, is given by

\begin{equation}\label{2.6}
Q(x ,\langle\delta\hat{x}^n\rangle) = V'(x,t) - \langle
V'(\hat{x},t)\rangle
\end{equation}

which by expressing  $ \hat{x}(t) = x(t) + \delta\hat{x}(t)$ in
$V(\hat{x},t)$ and using a Taylor series expansion around $x$ may
be rewritten as

\begin{equation}\label{2.7}
Q(x ,\langle\delta\hat{x}^n\rangle) =-\sum_{n\geq
2}\frac{1}{n!}V^{(n+1)}(x,t)\langle\delta\hat{x}^n\rangle
\end{equation}
Here $V^{(n+1)}$ the $(n+1)$th derivative with respect to $x$. The
calculation of $Q$ rests on the quantum correction terms
$\langle{\delta\hat{x}^n}\rangle$ which can be calculated order by
order by solving a set of quantum correction equations(as
discussed in the later part of this section). Furthermore the
quantum mechanical mean Langevin force is given by

\begin{equation}\label{2.8}
f(t) =
\sum_j\left[\langle\hat{q}_j(0)\rangle-\langle\hat{x}(0)\rangle
\kappa_j\cos\omega_jt +\kappa_j^{1/2}\hat{p}_j(0)
\sin\omega_jt\right]
\end{equation}

which must satisfy noise characteristics of the bath at
equilibrium ,

\begin{eqnarray}
\langle{f(t)} \rangle_S & = &
0\label{2.9}\\
\langle f(t) f(t') \rangle_S &=& \frac{1}{2} \sum_j \kappa_j\; \hbar
\omega_j \left( \coth \frac{\hbar \omega_j}{2 k T} \right) \cos
\omega_j (t-t')\nonumber\\\label{2.10}
\end{eqnarray}

Eq.(\ref{2.10}) expresses the quantum fluctuation-dissipation
relation. The above conditions Eq.(\ref{2.9})-Eq.(\ref {2.10}) can
be fulfilled provided the initial shifted co-ordinates
$\{\langle\hat{q}_j(0)\rangle-\langle\hat{x}(0)\rangle\}$ and
momenta $\langle{\hat{p}_j}(0)\rangle$ of the bath oscillators are
distributed according to the canonical thermal Wigner distribution
\cite{wig,hil} of the form

\begin{eqnarray}
&&P_j([\langle\hat{q}_j(0)\rangle-\langle\hat{x}(0)\rangle],
\langle\hat{p}_j(0)\rangle) \nonumber
\\
&=& N \exp\left\{-\;\frac{\frac {1}{2}\langle\hat{p}_j(0)\rangle^2 +
\frac
{1}{2}\kappa_j[\langle\hat{q}_j(0)\rangle-\langle\hat{x}(0)\rangle]^2}
{\hbar\omega_j[\overline{n}(\omega_j) +
\frac{1}{2}]}\right\}\nonumber \\\label{2.11}
\end{eqnarray}

so that the statistical averages $\langle...\rangle_s $ over the
quantum mechanical mean value $O_j$ of the bath variables are
defined as

\begin{equation}\label{2.12}
\langle O_j \rangle_s=\int O_j\; P_j\; d\langle
\hat{p}_j(0)\rangle\;
d\{\langle\hat{q}_j(0)\rangle-\langle\hat{x}(0)\rangle\}
\end{equation}

Here $\overline{n}(\omega)$ is given by Bose-Einstein
distributions $(e^{\frac{\hbar\omega}{kT}}-1)^{-1}$. $P_j$ is the
exact solution of Wigner equation for harmonic oscillator
\cite{wig,hil} and forms the basis for description of the quantum
noise characteristics of the bath kept in thermal equilibrium at
temperature $T$. $N$ is the normalization constant. In the
continuum limit the fluctuation-dissipation relation (\ref{2.10})
can be written as

\begin{eqnarray}
&&\langle {f(t)f(t')}\rangle_s\nonumber \\
&=&\frac{1}{2}\;\int_0^\infty d\omega\;
\kappa(\omega)\;\rho(\omega)\; \hbar\omega\;
\coth({\frac{\hbar\omega}{2kT}})\;\cos{\omega(t-t')}\nonumber
\\\label{2.13}
\end{eqnarray}

where we have introduced the density of the modes $\rho(\omega)$.
Since we are interested in the Markovian limit in the present
context, we assume  $\kappa(\omega)\rho(\omega)
=\frac{2}{\pi}\gamma$, and Eq.(\ref{2.13}) then yields

\begin{equation}\label{2.14}
\langle f(t)f(t')\rangle_s =2 D_q\delta(t-t')
\end{equation}

with

\begin{equation}\label{2.15}
D_q
=\frac{1}{2}\gamma\hbar\omega_0\coth{\frac{\hbar\omega_0}{2kT}}
\end{equation}
(The passage from Eq.(\ref{2.13}) to Eq.(\ref{2.14}) is given in the
appendix A)

 $\omega_0$ refers to the static frequency limit.
 Furthermore from Eq.(\ref{2.4}) in the continuum limit we have

\begin{equation}\label{2.16}
\gamma(t-t') = \gamma\;\delta(t-t')
\end{equation}

$\gamma$ is the dissipation constant in the Markovian limit. In
this limit Eq.(\ref{2.5}) therefore reduces to

\begin{equation}\label{2.17}
m \ddot{x} + \gamma \dot{x}+V'(x,t)= f(t ) + Q(x
,\langle\delta\hat{x}^n\rangle)
\end{equation}
In order to consider the activated processes in a bistable
potential with a fluctuating barrier height we consider the
potential of the form
\begin{equation}\label{2.18}
V(x,t) = U(x)+g(x)\xi(t),
\end{equation}
where $U(x)$ is a bistable potential
($-\frac{a}{2}x^2+\frac{b}{4}x^4$, $a$, $b$ being constants ) with
a barrier at the metastable point  $x=0$ and two stable points at
$x=\pm (\frac{a}{b})^{\frac{1}{2}}$. The fluctuations in the
potential are driven by an Ornstein-Uhlenbeck noise process

\begin{equation}\label{2.19}
\dot\xi(t)=-\frac{\xi(t)}{\tau}+\frac{\sqrt{2\sigma^2}}{\tau}\eta(t)
\end{equation}
where $\eta(t)$ is zero mean $\delta$-correlated Gaussian noise. The
stochastic process $\xi(t)$, can be characterized by the following
set of equations. The probability function $\bar{\rho}(\xi)$, the
variance $\sigma^2$ and the correlation function of the noise are
given
\begin{equation}\label{2.20}
\bar{\rho}(\xi)=\sqrt{2\pi\sigma^2}\exp(-\frac{\xi^2}{2\sigma^2})\;,
\end{equation}
where
\begin{equation}\label{2.21}
\sigma^2=\int_{-\infty}^{+\infty}\xi^2\bar{\rho}(\xi) d\xi\;,
\end{equation}
and
\begin{equation}\label{2.22}
\langle \xi(t)\xi(0)\rangle=\sigma^2\exp(-\frac{|t|}{\tau})\;,
\end{equation}
respectively. The barrier can also be subjected to dichotomic
fluctuation \cite{luc}, $\xi(t)=\{-\alpha,\beta\}$ that flips
between two values with flipping rate $\mu_{\alpha}$ and
$\mu_{\beta}$ respectively. This process can also be taylored as
zero mean valued, exponentially correlated and with a Gaussian
distribution.

\subsection{Overdamped limit}
To proceed further we now confine ourselves  to overdamped
condition so that the quantum Langevin equation in one variable
takes the following form:
\begin{equation}\label{2.23} \gamma \dot{x}=ax-bx^3+g'(x)
\xi(t)+f(t)+Q(x ,\langle\delta\hat{x}^n\rangle)
\end{equation}
The above classical-like stochastic differential equation contains
the essential quantum features, through the terms $f(t)$ which
represent the quantum noise of the heat bath and the another term
$Q(x ,\langle\delta\hat{x}^n\rangle)$ which essentially arises due
to nonlinear part of the system potential. The nonlinearity and the
quantum effects are entangled in the latter quantity modifying the
classical part of the potential. Thus the classical potential force
$-V'(x,t)$ is modified by the quantum dispersion. In absence of
quantum dispersion term $Q(x ,\langle\delta\hat{x}^n\rangle)$ and
with $D_q\rightarrow \gamma kT $ as one approaches the classical
limit ($kT\gg\hbar\omega_0$), the quantum Langevin equation reduces
to classical one. A probabilistic description of the system Eq.(\ref
{2.23}) is provided by the time-dependent probability distribution
function $\bar{P}(x,\xi,t)$. For a finite correlation time of the
noise driving the barrier to fast fluctuations it is
possible\cite{han} to go over to an approximate (exact in the limit
$\tau\rightarrow 0$) description which allows one to transform
externally driven Langevin equation Eq.(\ref {2.23}) into an
autonomous Langevin equation for the position variable $x$. The
corresponding equation for probability density function $P(x,t)$ is
given by

\begin{eqnarray}\label{2.24}
\frac{\partial P(x,t)}{\partial t}&=&-\frac{\partial}{\partial x}
\left[
\kappa(x,\tau)f(x)\right]P(x,t)\nonumber
\\
&+&\sigma^2\frac{\partial}{\partial
x} \left[ \kappa(x,\tau)g'(x)\frac{\partial}{\partial x}
\kappa(x,\tau)g'(x)\right]P(x,t)\nonumber
\\
&+&D_q\frac{\partial}{\partial x}
\left[ \kappa(x,\tau)\frac{\partial}{\partial x} \kappa(x,\tau)\right]P(x,t)
\end{eqnarray}

where $\gamma$ is assumed to unity, $\kappa(x,\tau)=\left[1-\tau
g'(x) \left(\frac{f(x)}{g'(x)} \right)'\right]^{-1}$ and
$f(x)=-U'(x)+Q(x ,\langle\delta\hat{x}^n\rangle)$. The quantum
nature of the above Fokker-Planck equation governing evolution of
$P(x,t)$ is manifested through the quantum correction to the
potential term and the quantum diffusion coefficient
characterizing the thermal bath.

The quantity of special interest here is the mean thermally
activated escape time of the particle from the well whose barrier is
subjected to fluctuation. The particle is governed by the stochastic
dynamics Eq.(\ref {2.24}). It is important to emphasize that the
typical mean escape time should be much larger than the time scale
of the deterministic quantum dynamics $\dot{x}=-U'(x)+Q(x
,\langle\delta\hat{x}^n\rangle)$ for nonzero fluctuation of the
potential well, and it is further required that the strengths of
thermal noise and barrier fluctuation must be small in comparison
with the barrier height. The solution of Eq.(\ref {2.24}) in the
stationary state reads as

\begin{eqnarray}\label{2.25}
&&P_s(x,\tau)= \frac{\left[1-\tau g'(x) \left(\frac{f(x)}{g'(x)}
\right)'\right]}{{D_q^{\frac{1}{2}}\left(1+\frac{\sigma^2}{D_q}g'(x)^2\right)}
^{\frac{1}{2}}}\nonumber
\\
&\times& \exp\left(\int_0^x f(y) \frac{\left[1-\tau g'(y)
\left(\frac{f(y)}{g'(y)}
\right)'\right]}{D_q\left(1+\frac{\sigma^2}{D_q}g'(y)^2\right)}dy
\right)
\end{eqnarray}
 The stationary probability distribution function $P_s(x,\tau)$ of the stochastic
 process $x(t)$, which is the quantum mechanical mean position, is bimodal
 in nature with nonzero probability current at the barrier top even at zero
  temperature characterizing the zero point contribution of the thermal bath
  (as $T\rightarrow 0$, $D_q \rightarrow \hbar\omega_0$). $P(x,\tau)$ is additionally
  modified over its classical nature by a quantum contribution characterizing
  a correction due to anharmonic part of the system potential.

To proceed further it is necessary to find out the quantum
correction term $Q(x ,\langle\delta\hat{x}^n\rangle)$ more
explicitly. To this end we return to the overdamped operator
equation (\ref{2.2}) and use $\hat{x}( t)=x( t)+
\delta\hat{x}(t)$, where $x( t)(=\langle\hat{x}(t)\rangle)$  is
the quantum mechanical mean value of the operator $\hat{x}$. By
construction $[\delta\hat{x},\delta\hat{p}]=i\hbar$ and
$\langle\delta \hat{x}\rangle=0$. We then obtain the quantum
correction equation in the overdamped limit after quantum
mechanical averaging with the coherent states over the bath
operators as

\begin{equation}\label{2.26}
\gamma\delta\dot{\hat{x}} +V''(x,t)\delta\hat{x}+ \sum_{n\geq
2}\frac{1}{n!}{V}^{n+1}(x,t)(\delta\hat{x}^n- \langle
\delta\hat{x}^n\rangle)=0
\end{equation}

With the help of operator equation (\ref{2.26}) we obtain the
equations for $\langle\delta\hat{x}^n\rangle$

\begin{equation}\label{2.27}
\frac{d}{dt}\langle\delta\hat{x}^2\rangle=-\frac{1}{\gamma}\left[2V''(x,t)\langle\delta\hat{x}^2\rangle
+V'''(x,t)\langle\delta\hat{x}^3\rangle\right]
\end{equation}

\begin{eqnarray}\label{2.28}
\frac{d}{dt}\langle\delta\hat{x}^3\rangle&=&-\frac{1}{\gamma}\left[
3V''(x,t)\langle\delta\hat{x}^3\rangle
+\frac{3}{2}V'''(x,t)\langle\delta\hat{x}^4\rangle\right]\nonumber
\\
&&+\left[\frac{3}{2 \gamma}V'''(x,t)
\langle\delta\hat{x}^2\rangle^2\right]
\end{eqnarray}
and so on. To take into account of the leading order contribution
$\langle\delta\hat{x}^2\rangle$ explicitly we may write
\begin{equation}\label{2.29}
d\langle\delta\hat{x}^2\rangle =
-\frac{2}{\gamma}V''(x,t)\langle\delta\hat{x}^2\rangle dt
\end{equation}
The overdamped deterministic motion on the other hand gives $\gamma
dx = -V'(x,t)dt$ which when used in Eq.(\ref {2.29}) yields after
integration
\begin{equation}\label{2.30}
\langle\delta\hat{x}^2\rangle=\Delta_q \left[ V'(x,t)\right]^2
\end{equation}
$\Delta_q$ is the quantum correction parameter, as given by
$\Delta_q=\frac{\langle\delta\hat{x}^2\rangle_{x_c}}{\left[
V'(x_c,t)\right]^2}$, $x_c$ being a given quantum mechanical mean
position.

\section{Resonant activation in the quantum system}

We are now in a position to analyze the quantum resonant activation.
In our present problem the fluctuating part of the potential
associated with harmonic term is assumed to be of the form
$g(x)=\frac{x^2}{2}$ and $x= \pm \sqrt{\frac{a}{b}}$ and $x=0$ are
the absolute minima and maximum of the potential $U(x)$
respectively. Having known the stationary distribution
Eq.(\ref{2.25}) along with the quantum correction in $f(x)$ the
calculation of stochastic dynamics is straightforward. This can be
obtained using the standard result on mean first passage time
\cite{str} as
\begin{equation}\label{3.1}
\langle T\rangle=\int^{0}_{-\sqrt{\frac{a}{b}}}
\frac{dx}{D(x,\tau)P_s(x)}\int_{-\infty}^x P_s(x)dy
\end{equation}
 where $D_{eff}(x,\tau)=D_q\left(1+Rg'(x)^2\right)\left[1-\tau g'(x)
\left(\frac{f(x)}{g'(x)} \right)'\right]^{-2}$
 to give
\begin{equation}\label{3.2}
\langle T\rangle=\frac{2\pi \sqrt{1+2\tau(a+\Delta_1)}}
{a\sqrt{2(1-\Delta_2)}}\exp\left[\frac{\Delta V_{eff}(R,\tau)}{D_q}\right]
\end{equation}

Here $\Delta_1$ and $\Delta_2$ both are the leading order quantum correction terms as given by
\begin{equation}\label{3.3}
\Delta_1=\left[-g'(x)\left(\frac{Q(x
,\langle\delta\hat{x}^n\rangle)}{g'(x)}\right)\right]_{x=-\sqrt{\frac{a}{b}}}
\end{equation}

\begin{equation}\label{3.4}
\Delta_2=\left[Q'(x
,\langle\delta\hat{x}^n\rangle)\right]_{x=-\sqrt{\frac{a}{b}}}
\end{equation}
and $R=\frac{\sigma^2}{D_q}$ ; $\Delta V_{eff}(R,\tau)$ is the
effective barrier height of the following form

\begin{eqnarray}
&&\Delta V_{eff}(R,\tau)\nonumber \\
&=& -\int^{0}_{-\sqrt{\frac{a}{b}}}\frac{\left( -U'(x)+Q\right)
\left[ 1+\tau
g'(x)\left(\frac{-U'(x)+Q}{g'(x)}\right)\right]dx}{\left(1+R
g'(x)^2\right)}\nonumber\\\label{3.5}
\end{eqnarray}
Eq.(\ref {3.2}) is the central analytical result of this paper. To
analyse the theoretical results on the essential features of the
activated escape following quantum stochastic dynamics over a
fluctuating potential well we now resort to numerical simulation of
the Eqs.(\ref {2.23}),(\ref{2.19}) and the quantum correction
equations (\ref {2.27}) and (\ref {2.28}) simultaneously using
standard Heun's algorithom.
 A very small time step ($\triangle t$) $0.001$ for
numerical integration has been used. In our simulation we follow the
dynamics of each particle starting in the left well at
$x=-\sqrt{\frac{a}{b}}$ till it arrives at the barrier top at $x=0$.
The first passage time being a statistical quantity due to the
random force we calculate the statistical average of the first
passage time over $1,000$ trajectories. We present the numerical
results in Fig.1 to Fig.4 for different parameters such as
temperatures, barrier height of the potential and noise strength of
the Ornstein-Unhenbeck noise process. All the curves exhibit minimum
at optimal $\tau$ values. Increasing strength of thermal or
non-thermal noise results in enhancement of escape rate and the
minima are shifted towards the origin. Physically this implies that
with increase in $D_q$ or $\sigma^2$, the escape time over the
barrier changes as a result of which the switching time of the
barrier fluctuations matches the order of escape time at lower
$\tau$ values. In order to examine the influence of the barrier
height on resonant activation we plot in Fig.3 the variation of mean
escape time as function of $\tau$ for several values of barrier
heights. Increase in the barrier height results in enhancement of
escape time and the minima are shifted to higher $\tau$ values.

We find that the barrier height Eq.(\ref {3.5}) monotonically
decreases to a limiting value
 with increasing $\tau$.
  Due to the presence of system
  nonlinearity the barrier height
 is slightly modified by an added slope. It is thus expected that
 during the temporary stay of the particle close to a minimum it
 makes unsuccessful attempts to escape and once the escape takes
 place it should occur at a slower rate in quantum case due to
 added slope to the effective potential than the corresponding
 classical case.
However, our theoretical and numerical calculations reveal that the
 mean escape time is lower in magnitude for the quantum particle at low
 temperature and the difference becomes insignificant at higher
 temperature. This is shown in the Fig.4. This behaviour can be
 interpreted in terms of an interplay between the quantum
 diffusion coefficient $D_q$ and the quantum correction due to
 system nonlinearity appearing in $V_{eff}(R,\tau)$.
When the temperature of the system is very low, i. e., in the vacuum
limit or in the deep tunneling region the anharmonic terms in the
potential do not contribute significantly. On the other hand as
temperature of the system increases significantly, $D_q$ increases
resulting in decrease of the effective potential and hence $D_q$ and
$Q$ compete to cancel the effect of each other at higher
temperature. Finally in Fig.5 we make a comparison between the
numerical simulation and the corresponding theoretical result. The
theoretical result agrees fairly well with our numerical result.

In addition to the barrier height the prefactor is also affected by
the quantum correction term. The approximate theoretical mean escape
time Eq.(\ref {3.2}) implies that both the barrier height and the
frequency factor have an important role to play with the activated
escape process.  We are now in a position to point out that the
thermally activated resonance phenomenon is controlled by three
different time scales as emphasized earlier \cite{rein}. The three
relavent time scales are the corresponding time of barrier
fluctuation ($0\ll\tau\ll \infty$), typical escape time $\hat T$,
which is much larger than the time scale describing deterministic
motion $\dot x=-U'(x)+Q'(x ,\langle\delta\hat{x}^n\rangle)$, the
third time scale being $T_a$ of the escape attempt ($T_a\ll \hat
T$).

\section{conclusion}
Based on the theoretical study and numerical simulation of quantum
stochastic dynamics in a double-well system with a fluctuating
barrier under influence of a Gaussian color noise, we have examined
the phenomenon of resonant activation. The governing equations are
classical looking in form but quantum mechanical in their content. A
key point of the present analysis is to describe the thermal bath in
terms of a canonical thermal Wigner distribution of harmonic
oscillators. This distribution remain positive definite even at
absolute zero signifying a pure state and allows us to look for the
external noise-induced resonance when the generic quantum effects in
the system make their presence felt. We have shown that quantization
significantly enhances resonant activation at low temperature due to
tunneling. For higher temperature as well as for stronger noise
strength the resonance effects get more pronounced. Since tunneling
accompanies activation, it is expected that the resonant activation
can be observed even at absolute zero, where the resonance effect
essentially is due to the vacuum field.

\begin{appendix}

\section{\textbf{(The passage from Eq.(\ref {2.13}) to Eq. (\ref {2.14}))}}
 We start from basic
definition(Louisell p-426) \cite{loui}
\begin{equation}\label{A1}
2D_q=\frac{1}{2\Delta{t}}\int_t^{t+\Delta{t}}dt\int_t^{t+\Delta{t}}dt'\;\;
\langle f(t)f(t')\rangle_s
\end{equation}
Using Eq.(\ref {2.13}) in Eq.(\ref{A1}) yields
\begin{eqnarray}\label{A2}
2D_q&=&\frac{1}{2\Delta{t}}\int_0^\infty
d\omega\;\kappa(\omega)\;\rho(\omega)\hbar\omega\;\coth\left(\frac{\hbar\omega}{2kT}\right)
\nonumber
\\
&&\times\int_t^{t+\Delta{t}}dt\int_t^{t+\Delta{t}}dt'\;\cos\omega(t-t')\nonumber\\\label{A2}
\end{eqnarray}
Explicit integration over time gives
\begin{equation}\label{A3}
2D_q=\frac{1}{2\Delta{t}}\int_0^\infty
d\omega\;\kappa(\omega)\;\rho(\omega)\hbar\omega\;\coth\left(\frac{\hbar\omega}{2kT}
\right)\;\;\;I(\omega,\Delta{t} )
\end{equation}
where
\begin{equation}\label{A4}
I(\omega,\;\Delta{t})=\frac{4}{w^2}\;\;\sin^2\frac{\omega\Delta{t}}{2}
\end{equation}
Putting $\kappa(\omega)\;\rho(\omega)=\frac{2}{\pi}\;\gamma$, we
obtain
\begin{equation}\label{A5}
 2D_q=\frac{\gamma}{\Delta{t}\;\pi}\int_0^\infty
d\omega\;\hbar\omega\;\coth\left(\frac{\hbar\omega}{2kT}\right)\;\;
\frac{\sin^2\frac{\omega\Delta{t}}{2}}{(\frac{\omega}{2})^2}
\end{equation}

Following Louisell(p $-426$) \cite{loui} we have under Markovian
condition, the correlation time $\tau_c\;\ll\;\Delta{t}$, the
coarse-grain time(over which the probability distribution function
evolves)

Thus as $\Delta{t}\rightarrow \infty$(in scale of $\tau_c$ which
goes to zero) the function
$\frac{\sin^2\frac{\omega\Delta{t}}{2}}{(\frac{\omega}{2})^2}$
oscillates violently so that one takes the slowly varying quantity
$[\hbar\omega\;\coth\frac{\hbar\omega}{2kT}]$ out of the integration
over frequency with an average value
$\hbar\omega_0\coth\frac{\hbar\omega_0}{2kT}$, $\omega_0$ be an
average static frequency. Since the integral
$\int_\infty^\infty\frac{\sin^2x\Delta{t}}{x^2}dx=\pi\Delta{t}$ it
follows immediately from Eq.(\ref{A5})

\begin{equation}\label{A6}
2 D_q=\gamma\; \hbar \omega_0 \coth\left(\frac{\hbar \omega_0}{2 k
T}\right)
\end{equation}

as given in Eq.(\ref{2.15}).

Again starting from Eq.(\ref{2.13}), we use the same argument as
before to have

\begin{eqnarray}\nonumber
&&\langle f(t)f(t')\rangle_s
\nonumber \\
&&= \frac{1}{2}\int_0^\infty d\omega\; \kappa(\omega)\rho(\omega)
\hbar \omega \coth\left(\frac{\hbar \omega}{2 k T}\right)
\cos\omega(t-t')\nonumber
\end{eqnarray}

and we use

\begin{eqnarray}
\int_0^\infty d\omega \cos \omega \tau = \pi\; \delta(\tau)\nonumber
\end{eqnarray}

to obtain

\begin{eqnarray}
&&\langle f(t)f(t') \rangle_s\nonumber \\
 &&= \frac{1}{2}
\int_0^\infty d\omega \left[\frac{2}{\pi} \gamma \right] \hbar
\omega \coth\left(\frac{\hbar \omega}{2 k
T}\right) \cos\omega(t-t')\nonumber\\
&&= \frac{\gamma\;\hbar \omega_0}{\pi}\coth\left(\frac{\hbar
\omega_0}{2 k T}\right) \pi \delta(t-t')\nonumber\\
&&=\gamma\; \hbar \omega_0 \coth\left(\frac{\hbar \omega_0}{2 k
T}\right) \delta(t-t')\label{A7}
\end{eqnarray}

Therefore from Eq.(\ref{A6}) and Eq.(\ref{A7}) we have

\begin{eqnarray}
\langle f(t)f(t') \rangle_s &=& 2D_q \delta(t-t')\nonumber
\end{eqnarray}

which is Eq.(\ref{2.14}).

Thus the derivation within Markovian approximation clearly depends
on the time scale separation. The results are valid even at absolute
zero as emphasized by Louisell.

\end{appendix}

\acknowledgments Thanks are due to the Council of Scientific and
industrial research, Govt. of India, for partial financial
support.

\begin{center}
{\bf Figure Captions}
\end{center}

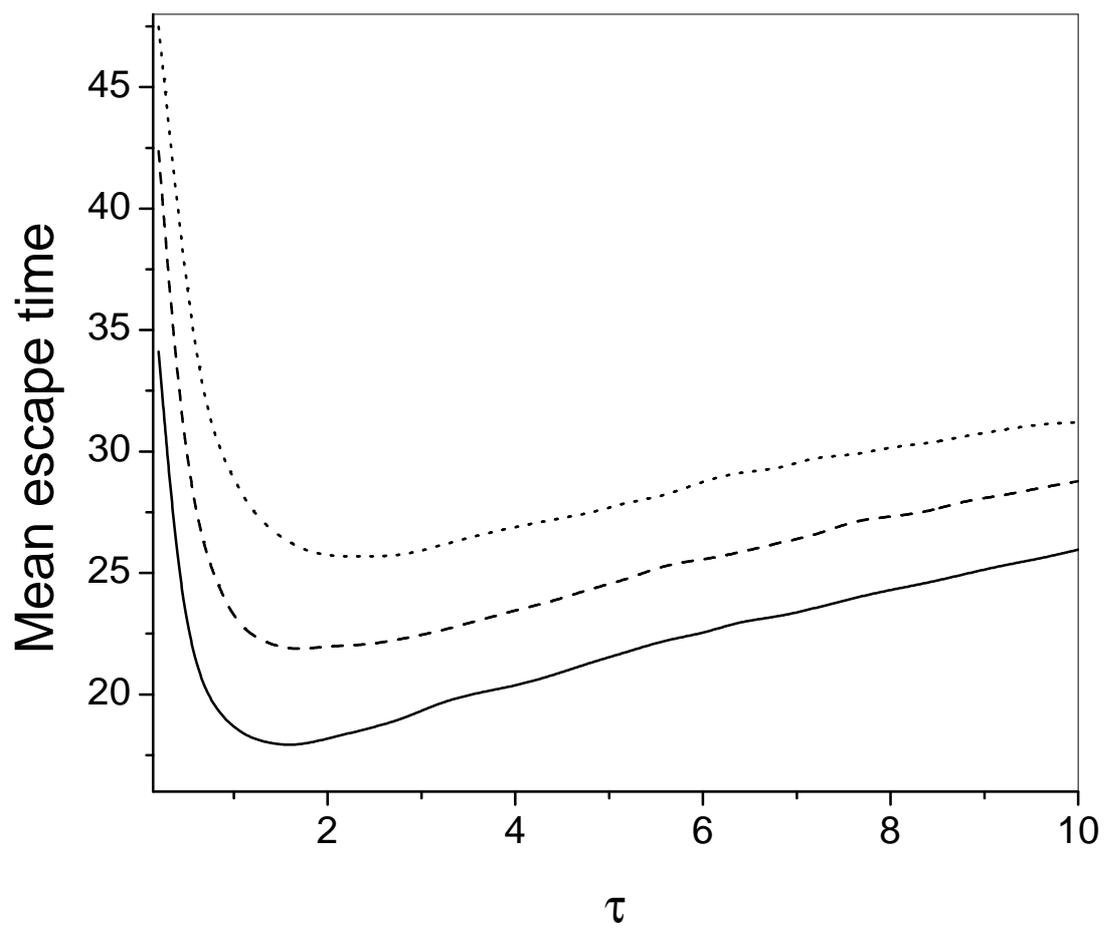
Fig.1. A plot of mean escape time vs $\tau$ using Gaussian colour
noise for different temperatures [(i) $T=0.1$ (dotted line), (ii)
$T=1.0$  (dashed line), (iii) $T=2.0$ (solid line)] and the other
parameter set $a=0.5$,
 $b=0.005$ and $\sigma^2=1.0$.

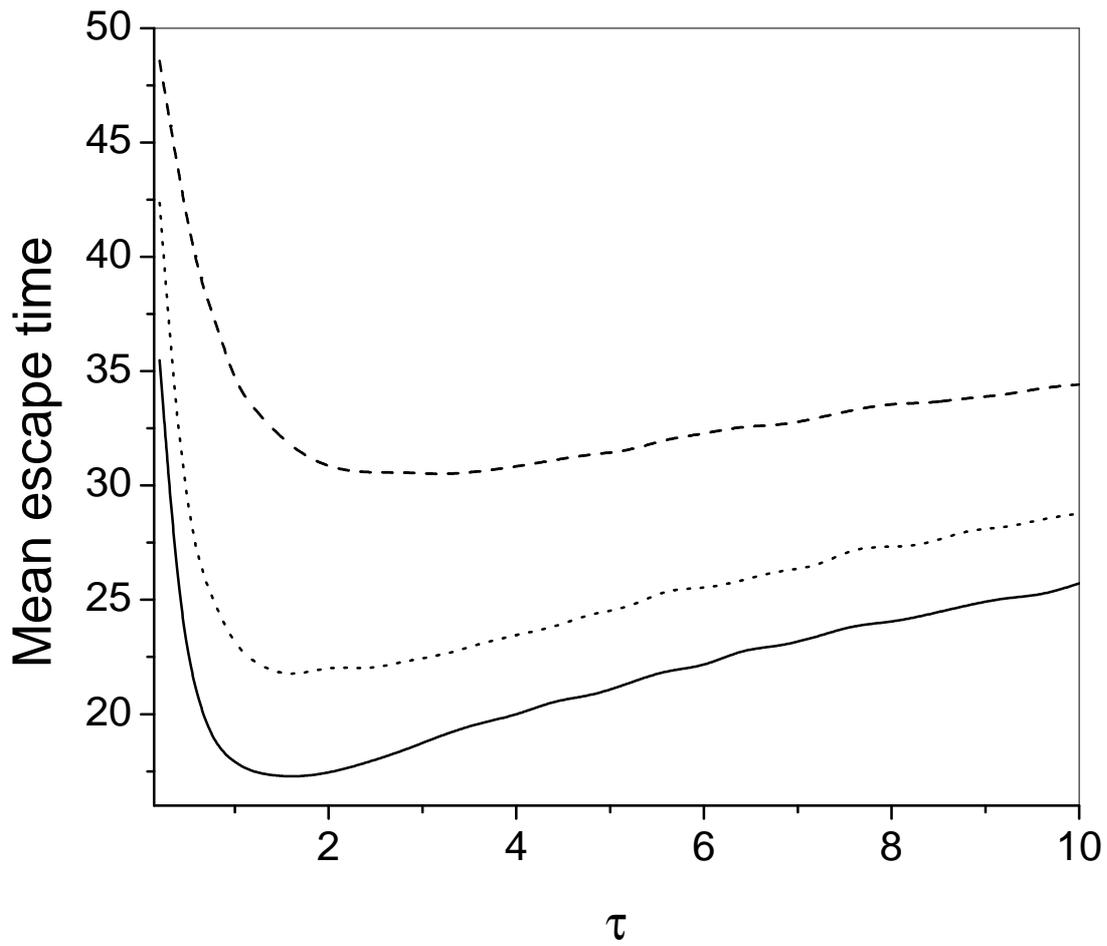
Fig.2. A plot of mean escape time vs $\tau$ using Gaussian colour
noise for different $\sigma^2$ [(i)$\sigma^2=0.5$ (dashed line),
(ii) $\sigma^2=1.0$  (dotted line), (iii) $\sigma^2=1.5$ (solid
line)] and the other parameter
 set $a=0.5$, $b=0.005$ and $T=1.0$.

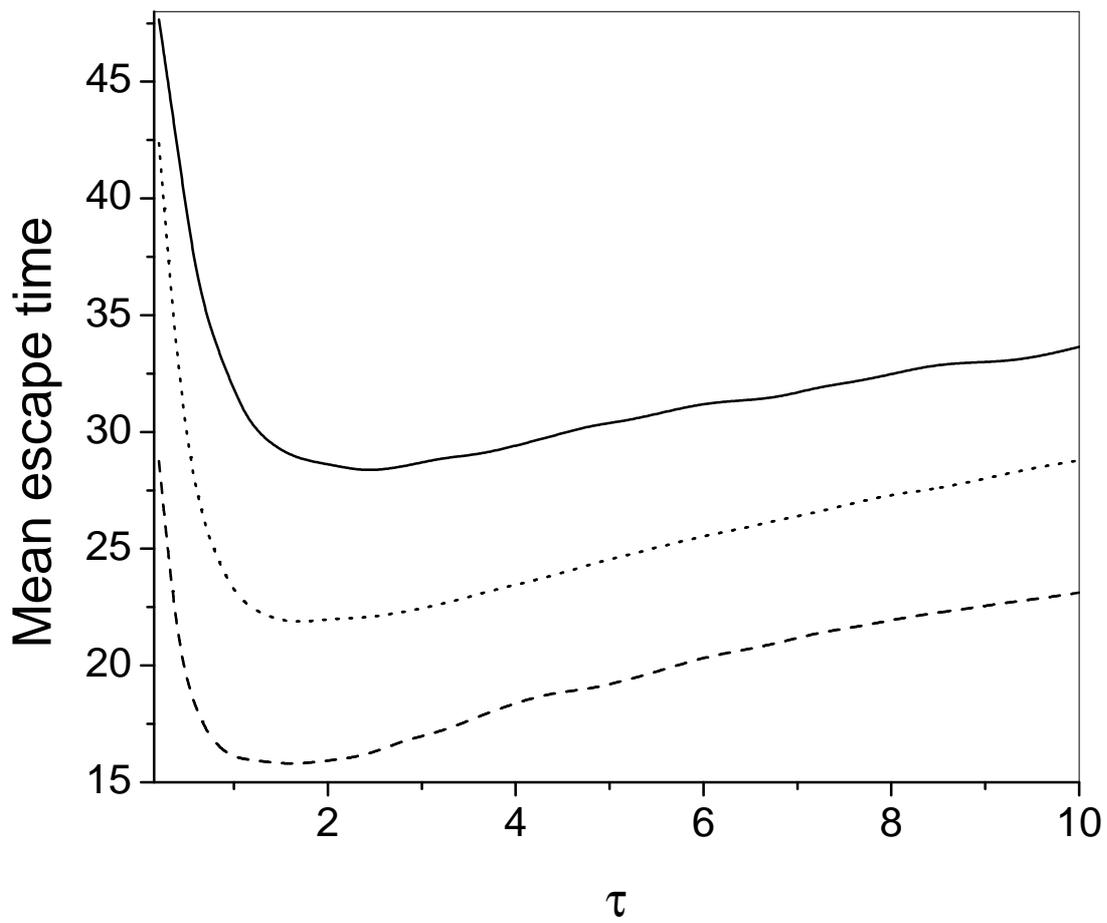
Fig.3. A plot of mean escape time vs $\tau$ using Gaussian colour
noise for different barrier heights [(i) $a=0.3$ (dashed line), (ii)
$a=0.5$  (dotted line), (iii) $a=0.7$ (solid line)]
 and the other parameter set $b=0.005$, $\sigma^2=1.0$ and
 $T=1.0$.

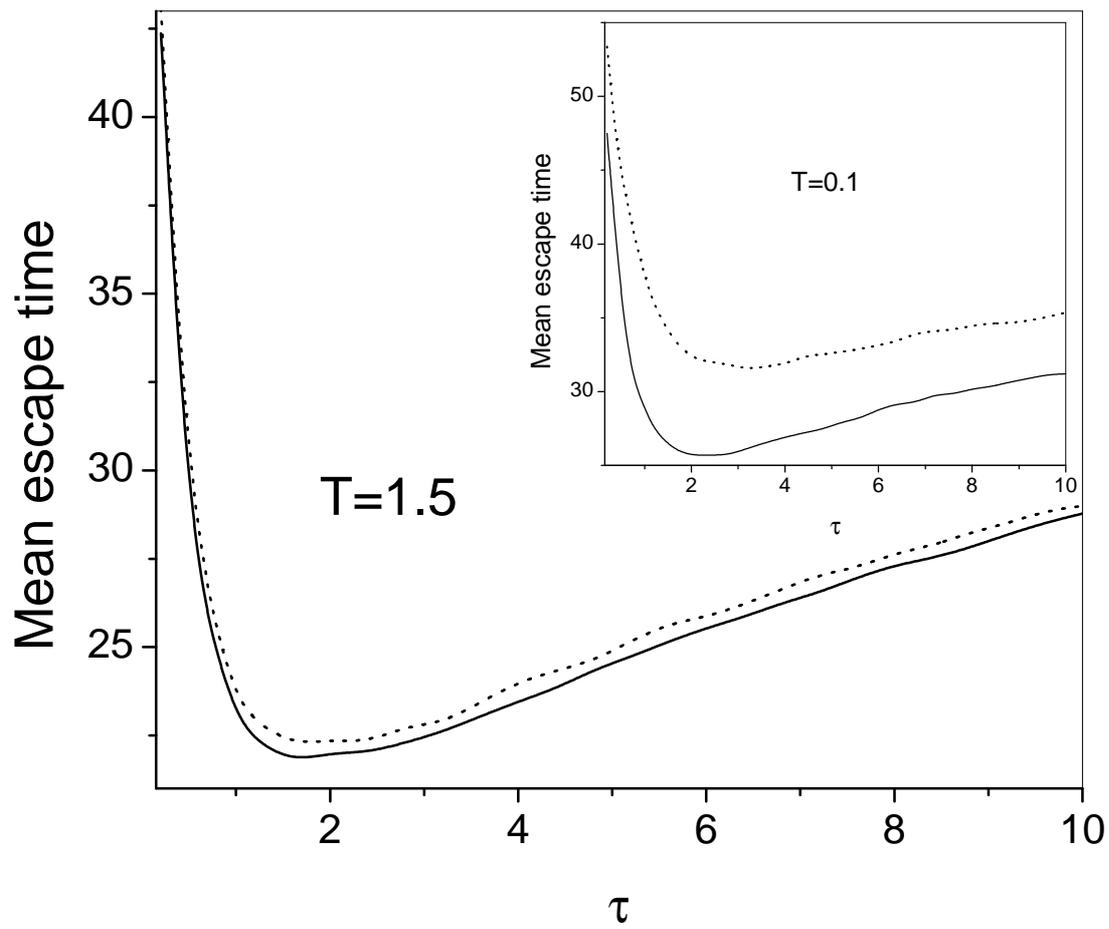
Fig.4. A comparison between classical(dotted line) and quantum
(solid line) mean escape time describing resonant activation
phenomenon using Gaussian colour noise for the parameter set
$a=0.5$, $b=0.005$ and $\sigma^2=1.0$ at two different temperatures
$T=1.5$ and $T=0.1$.

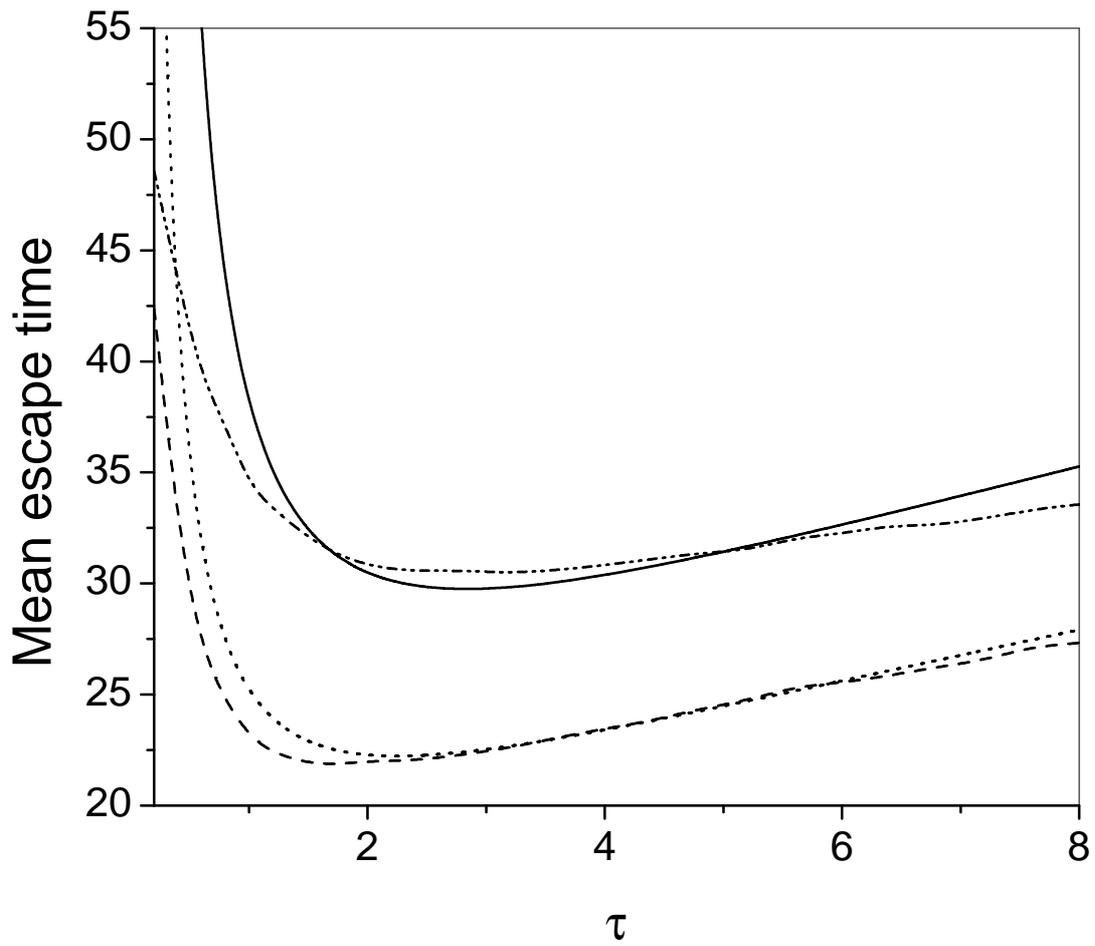
Fig.5. A comparison between numerical(dashed dot dot line and dashed
line) and analytical (solid and dotted line) results using Gaussian
colour noise for different $\sigma^2$ [(i) $\sigma^2=0.5)$ (solid
and dashed dot dot line), (ii) $\sigma^2=1.5$ (dotted and dashed
line)] and the other parameter set $a=0.5$, $b=0.005$ and $T=1.5$.

\end{document}